# Method for processing and analyzing tectonic earthquake aftershocks: A new look at old problem

A.V. Guglielmi

*Schmidt Institute of Physics of the Earth, Russian Academy of Sciences; Bol'shaya Gruzinskaya str., 10, bld. 1, Moscow, 123242 Russia; guglielmi@mail.ru*

**Abstract**. The deactivation coefficient has been introduced for the source that "cools down" after the mainshock of the earthquake. The deactivation coefficient satisfies two conditions. First, it can be computed from observed aftershock frequency data (*axiom of computability*). Second, the deactivation factor is independent of time if and only if Omori's law is strictly satisfied (*hyperbolicity axiom*). From the two axioms follows an explicit expression for calculating the source deactivation factor from observed aftershock data. A simple and efficient method for processing and analyzing aftershocks using the deactivation factor has been developed. A two-stage mode of the source relaxation to a new equilibrium state after the formation of the main rupture of rock continuity in it was discovered. In the first stage, called the *Omori epoch*, the Omori law is strictly followed. The duration of the first stage is tentatively 10 to 100 days. In the Omori epoch, the average frequency of aftershocks is predictable with high accuracy. At the end of the Omori epoch, a bifurcation occurs. In the second stage, the deactivation factor changes with time in an unpredictable way. The phase portrait of the source allows us to visualize the complex dynamics of seismic activity in the second stage of the source relaxation. The experiment shows that the Hirano-Utsu law is inapplicable for describing the evolution of aftershocks neither at the first nor at the second stage of the source relaxation.

*Keywords*: deactivation factor, evolution equation, phase portrait, Omori's law, Hirano-Utsu's law.

## 1. Introduction

In the late nineteenth century, Milne created a horizontal pendulum seismometer, and his student Omori, after processing and analyzing seismometric data, formulated historically the first law of earthquake physics [1, 2]. Omori's law states that the average aftershock frequency $n$ decreases hyperbolically with time:

$$n(t) = \frac{k}{c+t}. \qquad (1)$$

Here $k > 0$, $c > 0$, $t \geq 0$ [3].

The simplicity and elegance of Omori's law is very attractive. In the 130 years since then, the law has undergone only one radical modification. Namely, 100 years ago Hirano [4] changed the law, presenting it in the form

$$n(t) = \frac{k}{(c+t)^p}, \qquad (2)$$

where $p > 0$. In 1938, the Hirano formula was mentioned by Jeffries [5], but after that it was forgotten for many years and the reason for this was the tragic events that were then happening in the world. Only in the second half of the last century did formula (2), figuratively speaking, gain a second wind thanks to Utsu's research [6–8]. From then to the present day, formula (2) has been widely used in the processing and analysis of aftershocks. The presentation of the original theory of ETAS (*Epidemic type aftershock sequence*) begins with the Hirano formula [9]. Formula (2) is sometimes called the "Modified Omori function" [10], sometimes the "Omori–Utsu law" [11], and sometimes the "Utsu law" [12]. I think it would be correct to call (2) the Hirano-Utsu formula.

Processing observations using the Hirano-Utsu formula gives us two numerical parameters, $k$ and $p$, characterizing the source of the earthquake. It turns out that the dimension of the parameter $k$ depends on the value of the parameter $p$. For example, at $p = 1.2$ we have $[k] = s^{+0.2}$, and at $p = 0.9$ we get another dimension: $[k] = s^{-0.1}$. This is not entirely satisfactory, since it is generally accepted that the phenomenological parameters characterizing a physical system have a fixed dimension. Note that in this respect Omori formula (1) is quite correct.

We have the deepest respect for the outstanding achievements of the founders of modern seismology. However, it is quite clear that over time the research methodology should be enriched and improved. Guided by this consideration, a small team of geophysicists from the Institute of Earth Physics of the Russian Academy of Sciences has made efforts in recent years to create a method for processing and analyzing aftershocks, different from the Hirano-Utsu method we are used to. The general principles of the new approach to the study of aftershocks are described in papers [13–23]. The new method is simple in practical application and quite effective. But so far the new methodology does not appear to have been fully understood and, moreover, has sometimes been subject to criticism. The purpose of this paper is to explain the new procedure for processing and analyzing aftershocks in a very simple way and to illustrate the effectiveness of our method with clear examples.

## 2. Source deactivation coefficient

As the subject of research, we choose the source of the earthquake, and not the aftershocks themselves. The state of the source after the formation of a main rupture in it will be described phenomenologically by the parameter $\sigma(t)$, which we will call the source deactivation coefficient [2, 13]. The deactivation coefficient generally characterizes the current ability of the source to excite aftershocks. The dependence of the deactivation coefficient on time reflects the nonstationarity of the source, relaxing to a new state of equilibrium after the main rupture of rock continuity, which manifested itself in the form of the main shock of the earthquake.

We impose two conditions on the value of the deactivation coefficient. The first condition (*computability axiom*) is that the deactivation coefficient $\sigma(t)$ can be calculated from observation data of the aftershock frequency $n(t)$. The second condition (*hyperbolicity axiom*) is more specific: we require that $\sigma = \text{const}$ if and only if Omori's law is strictly satisfied. From our axioms it follows that

$$\sigma(t) = -\frac{1}{n^2(t)} \frac{dn(t)}{dt}. \qquad (3)$$

In fact, let's make a replacement of the dependent variable: $g = 1/n$. It can be seen that (3) is equivalent to the linear differential equation for the auxiliary function $g(t)$:

$$\dot{g} = \sigma. \qquad (4)$$

Here the dot above the symbol means differentiation with respect to time. We see that our axiomatics leads to an equation for the evolution of aftershocks in the form of the simplest differential equation. The solution to the equation has the form

$$n(t) = \frac{n_0}{1 + n_0 \tau(t)}, \tag{5}$$

where $n_0 = n(0)$ is the initial condition, and $\tau(t)$ is the so-called proper time of the source, equal to

$$\tau(t) = \int_0^t \sigma(t')dt'. \tag{6}$$

For $\sigma = \text{const}$ we have

$$n(t) = \frac{n_0}{1 + n_0 \sigma t}, \tag{7}$$

which coincides with Omori's law (1) up to notation. Thus, it is quite obvious that formula (3) satisfies the hyperbolicity axiom.

To practically satisfy the computability axiom, a little extra work needs to be done. Let us pose an inverse problem for the evolution equation: determine the deactivation coefficient based on aftershock observation data. The trivial solution $\sigma = \dot{g}$ turns out to be unstable due to rapid fluctuations of the auxiliary function. In other words, the inverse problem, as often happens, is formulated incorrectly. Regularization in this case comes down to smoothing the auxiliary function. As a result, the solution takes the form

$$\sigma(t) = \frac{d}{dt}\langle g(t) \rangle, \tag{8}$$

where the angle brackets indicate the smoothing procedure.

## 3. Two-stage relaxation of the earthquake source

So, processing aftershocks using the new method is quite simple and consists of replacing $n(t) \to g(t)$, smoothing the auxiliary function $g(t)$, and calculating $\sigma(t)$ using formula (8). Based on the processing results, an Atlas of Aftershocks was compiled [14], containing information on the deactivation coefficient $\sigma(t)$ for several dozen events. The analysis of the Atlas led to the following important conclusions.

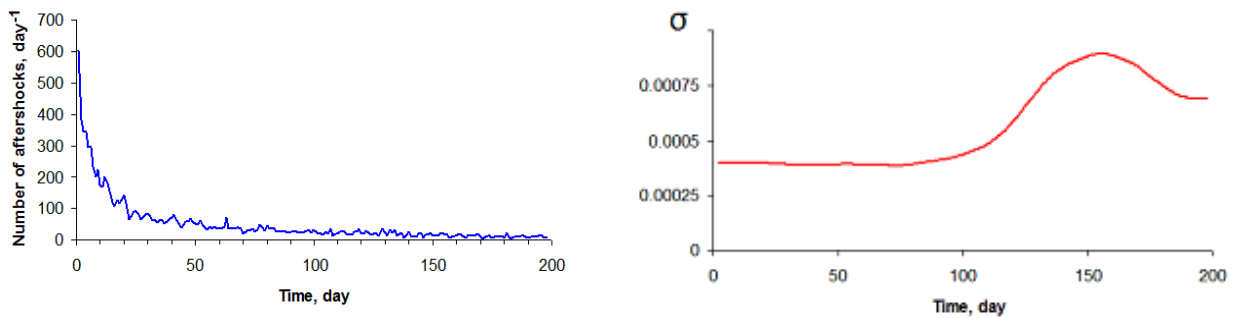

**Fig. 1**. An example of aftershocks with a very long Omori epoch. The main sh0ck with magnitude $M$ = 6.7 occurred on 1994.01.17 in Southern California at a depth of $H$ = 18 km [14].

The evolution of the source is divided into two stages. Two-stage evolution was clearly observed in all cases considered. The two stages differ sharply in the nature of the time dependence of the deactivation coefficient. At the first stage $\sigma = \text{const}$. In other words, Omori law is strictly observed. We called this stage the *Omori epoch*. The duration of the Omori epoch varies from case to case from approximately 10 to 100 days. In Figure 1 we see the very long Omori epoch.

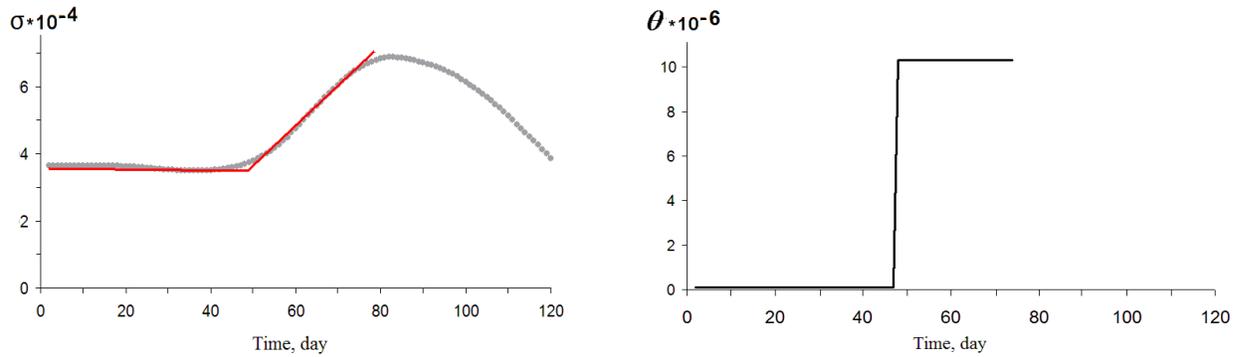

**Fig. 2**. Deactivation coefficient (left) and its piecewise linear approximation. The time derivative of the deactivation coefficient is shown on the right. An earthquake occurred in Northern California on July 20, 1986. The hypocenter depth is 6 km, $M = 5.9$ [19].

Thus, at the first stage of relaxation $\sigma = \text{const}$ and the average frequency of aftershocks changes with time in a completely predictable manner. But the Omori epoch ends abruptly, the bifurcation of the source occurs, and the second stage begins. The deactivation factor changes unpredictably over time. Figure 2 clearly illustrates the phenomenon of bifurcation. The Omori epoch in this case ends 49 days after the main shock. From the beginning of the second stage, the deactivation coefficient increases sharply. (Note that the decrease in $\sigma(t)$ after bmfurcation is just as typical.) In the right panel of the figure we see a characteristic jump in the function $\theta = d\sigma/dt$ at the moment of bifurcation. It should be especially emphasized that at the second stage of the evolution of the source, the function $\sigma(t)$ is non-monotonic. The activity of aftershocks in the second stage changes over time in unpredictable ways. This is in sharp contrast to the predictability of the average frequency of aftershocks in the first stage. An analogy arises with the transition of a hydrodynamic flow from a laminar to a turbulent flow regime.

As a result of processing aftershocks and analyzing the deactivation coefficient, three numbers are obtained: the deactivation coefficient $\sigma_0$ at the Omori epoch, the duration of the Omori epoch $\Delta t$, and the magnitude of the jump $\Delta\theta$ at the moment of bifurcation. This additional information can be used to study the properties of the earthquake source.

## 4. Phase portrait of the source

Let us give another example of using the deactivation coefficient to study the dynamics of the source. We will talk about constructing a phase portrait of the source.

The idea of constructing a phase portrait of a dynamic system (4) belongs to Faraoni [24]. It should be taken into account that we describe the evolution of the source in an extremely simplified manner. We do not yet have the ability to represent the phase trajectory in the multidimensional phase space of the source. All that remains for us is to look for a suitable section of the phase space by the phase plane and find the projection of the trajectory onto the subspace chosen in this way. Faraoni presented the projection of the trajectory onto the phase half-plane $(n,\dot{n})$, $n \geq 0$. The phase portrait consists of a single trajectory, which has the form of a semi-parabola with the vertex at point $(0,0)$, which is the point of stable equilibrium. Faraoni's technique was used in [15] to construct a family of phase trajectories of the logistic equation that simulates the evolution of aftershocks.

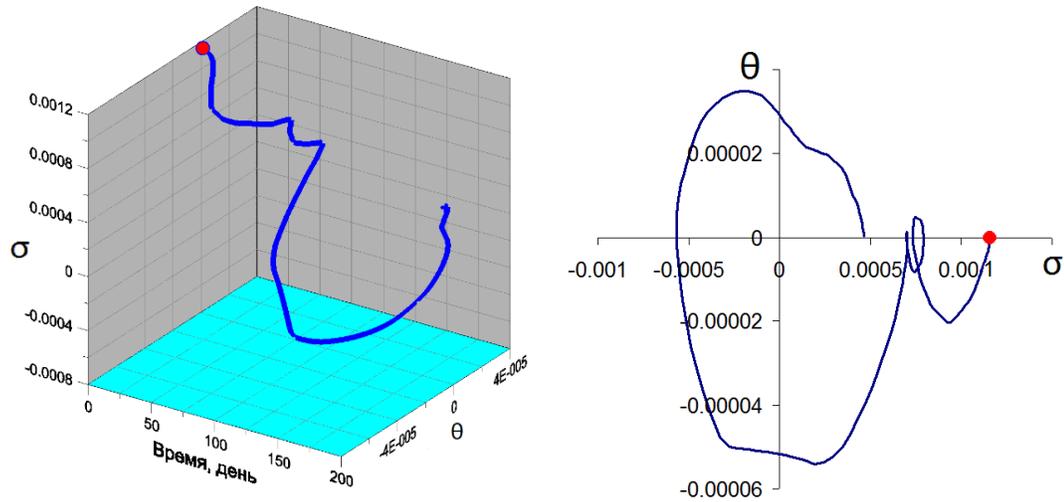

**Fig. 3**. Graph of the movement of the representing point in coordinates $\sigma(t)$, $\theta(t)$ (left) and the corresponding phase portrait (right). An earthquake occurred in Northern California on January 7, 1983. The magnitude of the main shock is $M = 5.4$. The red dot marks the beginning of the phase trajectory.

However, it was not possible to identify the similarity between theoretical and real phase portraits of the source in the phase plane $(n, \dot{n})$. Therefore, it was decided to search for the source image experimentally in the phase plane $(\sigma, \theta)$ and in the extended phase plane $(\sigma, \theta, t)$. In Figure 3 we see a typical example of a rather complex movement of a representing point. Our research group plans to study phase portraits of this kind. It seems to us that the approach to studying the dynamics of the source, based on the study of phase portraits, leads to effective results. Ideally, we would like to use phase portraits to find dynamic equations that sufficiently fully describe the evolution of the source.

## 5. Discussion

We imagine the source of an earthquake as a stressed-strained rock mass in which a major discontinuity spontaneously occurred, which manifested itself in the form of a main shock. The main shock is usually, but not always [17, 21], followed by a long series of aftershocks. The boundary of the source can be considered the convex hull of the minimum enveloping surface of all aftershocks.

We do not know for sure the dynamics of the source. We are dealing with a kind of "black box". The author, being a radiophysicist by training, is well acquainted with the classical method of searching for dynamic equations that simulate the behavior of a black box through experiments using the "input-output" scheme. The method works great in the case of linear dynamic systems, but the source is definitely a nonlinear system. In addition to this, with rare exceptions, we do not have input signals with which to probe the source. (The exceptions mentioned are, for example, the round-the-world seismic echo and free Earth vibrations, which can be used as pulsed and periodic input signals, respectively [25].) Thus, we are dealing with a *black box without an input*. We have to study the dynamics of the source from the output signal alone, for example, from the observed frequency of aftershocks $n(t)$.

Gudzenko [26] developed an original method for studying self-oscillating systems using only one output signal. It seems to us that Gudzenko's method can be modified and applied to study the nonlinear dynamics of a source relaxing to a new equilibrium state after the main shock of an earthquake, but this requires complex additional work to analyze aftershock fluctuations. In the review [27], when studying the dynamics of the source, the general ideas of the mathematical theory of disasters were used. The authors of a series of studies [13–23] were guided by the general principles of Husserl's phenomenology [28]. In this way, the methodology for extracting information about the properties of the earthquake source is gradually enriched.

This paper shows how to approach the problem of processing and analyzing aftershocks axiomatically. The two simple axioms that we have chosen as self-evident statements are independent, consistent and, what is surprising and pleasing, turned out to be quite complete in the sense that the consequences from them led to a number of non-trivial discoveries. The discovery of a two-stage source relaxation regime, the strict fulfillment of Omori's law at the first stage, the phenomenon of bifurcation and the complete unpredictability of dynamics at the second stage seem to us to be remarkable results. The ability to visualize the dynamics of the source at the second stage in the form of a phase portrait of the source also deserves attention. In general, the new method has certainly enriched the tools for studying earthquakes.

In connection with the linear equation for the evolution of the source (4), it is appropriate to make one clarification. Instead of (4), work [13] considers the simplest nonlinear evolution equation

$$\dot{n} + \sigma n^2 = 0, \qquad (9)$$

but this does not change the physics. Only the mathematical representation of the same reality changes. The fundamental innovation in the linear equation (4) and in the nonlinear equation (9) is the dependence of the deactivation coefficient on time. This is what allowed us to experimentally discover a two-stage mode of source relaxation. If $\sigma = \text{const}$, then according to the second axiom, Omori's law is satisfied. Our discovery is that Omori's law is strictly satisfied, but only at the first stage of the source evolution, which lasts from several days to several months and ends with bifurcation. In honor of Fusakichi Omori, we called the first stage the Omori epoch. During the Omori epoch, aftershock activity is quite predictable. In the second stage, the evolution is unpredictable, since the variation of the deactivation coefficient is chaotic. Thus, using our research method experimentally, we encountered a mysterious pattern of transition from a predictable stage to an unpredictable stage in the evolution of aftershocks. We have neither a dynamic equation nor even a simple fitting formula for the second stage. But at least we have found a phase portrait of the source, allowing us to visualize the strange dynamics of aftershocks at the second stage of relaxation.

Our axiomatics are incompatible with the views of Hirano [4], who considered the Omori formula (1) as a primitive fitting formula. He noticed that the Omori formula (1) does not satisfactorily approximate the evolution of aftershocks as an integral process. It would seem that from this observation it clearly follows that it is very likely that there is no simple algebraic formula that describes the dynamics of aftershocks holistically. But, contrary to the obvious, Hirano tried to "improve" the Omori law by introducing an additional fitting parameter $p$, and this, as one would expect, reduced the deviation of the experimental points from the fitting function (2). But it was possible to make other changes to the classical Omori law, for example, using series of functions with the inevitable introduction of new additional parameters to reduce the scatter of points. This would further hinder the ability to extract useful geophysical information from the observations.

It is quite clear that the empirical selection of simple formulas for approximating a set of experimental points is acceptable and is widely used in practice. Information that is hidden in the original data is extracted in this way and presented to the experimenter in the form of a small set of numbers. Processing aftershocks using the Omori formula (1) gives us the parameter $k$, which characterizes this particular source. If we process the observation data using the Hirano formula (2), then we obtain two parameters, $k$ and $p$. With his careful research, Utsu attracted the attention of seismologists to the possibility of extracting quantitative information about aftershocks in the form of a pair of numbers $k$ and $p$ (see, for example, the review [8]).

We are sympathetic to Hirano and Utsu search for methods for processing experimental data, but we do not want to lose sight of the fact that the relaxation of the source after the formation of a main rupture in it proceeds unevenly, and even unpredictably. In this case, it is impossible to describe relaxation holistically with a simple formula like (2). This is what prompted us to develop a methodology for processing and analyzing aftershocks, free from a priori considerations about the form of the law of decrease in the frequency of aftershocks over time, be it Omori's law or the Hirano-Utsu law. As a result, we discovered two stages of source relaxation, and at the first stage and only at the first the Omori law is valid, while the Hirano-Utsu law is not observed at either the first or the second stage of relaxation. In my opinion, the inapplicability of the Hirano-Utsu formula (2) for a holistic representation of aftershocks indicates that it would be a big mistake to take our proposed new method for studying aftershocks lightly.

A specialist in applied aspects of geophysics may have a question: what, in fact, does the use of a new method of processing and analyzing aftershocks give us in practice? It must be said frankly that so far our approach gives practically nothing, if we talk, for example, about predicting earthquakes. One could mention a probabilistic forecast of the strongest aftershock [25, 29], or a deterministic forecast of the frequency of aftershocks at the first stage of evolution, but the central question of the place, time and magnitude of the main shock remains uncertain (see [30], where the problem of forecast is discussed in more details).

## 6. Conclusion

Our approach to processing observations is based on simple and natural considerations. The object of study is the dynamics of the earthquake source relaxing after the main shock. The state of the source is described by the deactivation coefficient, which satisfies two independent, consistent and fairly complete axioms. From the axioms follows a simple expression for calculating the deactivation coefficient (3), equivalent to the simplest differential evolution equation (4). For the evolution equation, an inverse problem is posed, which turns out to be incorrect, but is easily regularized by smoothing an auxiliary function known from observations. The result of the processing is the deactivation coefficient, which flexibly describes the relaxation of the source as a holistic process. Analysis of the deactivation coefficient led to the discovery of previously unknown essential properties of the source.

We conclude the paper by listing some results obtained by our method of processing and analyzing aftershocks (for more details see [13–23]):

– With apodictic obviousness, there are two stages of source relaxation that differ sharply in the nature of the time dependence of the deactivation coefficient.

– At the first stage, Omori's law is strictly respected, as evidenced by the constancy of the deactivation coefficient.

– At the second stage, the deactivation coefficient is highly variable, its behavior over time becomes unpredictable.

– A hypothesis has been proposed about the bifurcation of the source at the end of the first stage of relaxation.

– The Hirano-Utsu law, which was previously used to process and analyze aftershocks, is not fulfilled either at the first or at the second stage of source relaxation.

– A classification of the so-called earthquake triads was made, and six types of triads were identified, differing from each other in clearly defined characteristics.

– In addition to Bath's law, which indicates the maximum magnitude of aftershocks, Zotov's law has been established, which indicates the maximum magnitude of foreshocks in mirror triads.

– Two previously unknown endogenous triggers of aftershocks were discovered, one of which is impulsive (round-the-world seismic echo), and the other periodic (free oscillations of the Earth).

*Acknowledgements.* I have been working on the topic of this paper for many years as part of a small team of geophysicists together with B.I. Klain, A.D. Zavlyalov and O.D. Zotov. To all of them I express my deep gratitude. This article is written based on the results of our collaborative research. For the interest in the work of our creative team and unwavering professional support, I sincerely thank A.L. Buchachenko, F.Z. Feygin, A.S. Potapov, and M.V. Rodkin. The work was carried out according to the plan of state assignments of Schmidt Institute of Physics of the Earth, Russian Academy of Sciences.